# Impact of Co$_2$C Nanoparticles on Enhancing the Critical Current Density of Bi-2223 Superconductor

Md. Arif Ali[1], Sourav M Karan[1], Nirmal Roy[1], S. S. Banerjee[1,*]

[1]Department of Physics, Indian Institute of Technology Kanpur, Kanpur, Uttar Pradesh 208016, India

*Corresponding author email ID : satyajit@iitk.ac.in

**Abstract:** We have investigated the superconducting properties of nanocomposite pellets made from Bi-2223 and Co$_2$C powders. Our measurements reveal loss of superconducting fraction in the nanocomposites, however, the retained superconducting fraction exhibits robust bulk superconducting properties. The $T_c$ of the retained superconducting fraction was 109 K which was found to be comparable to that of the pure Bi-2223 pellet. We found that the composites net magnetization response is a superposition of ferromagnetic and superconducting fractions contributions. Analysis revealed that the surviving superconducting fraction exhibits a robust Meissner response. In the nanocomposite the irreversibility field of the superconducting fraction at 77 K is found to increase by almost three times compared to the pristine material, thereby showing strong vortex pinning features. We also find a broadened magnetic field regime over which we observe a single vortex pinning regime sustained in the nanocomposite. The critical current density, $J_c$, of the nanocomposite was found to be approximately five times higher than that of the pristine Bi-2223 pellet at low $T$. In fact, the enhancement in $J_c$ is most significant in the high $T$ regime, where at temperatures close to $T_c$ in the nanocomposite we see almost two orders of magnitude increase of $J_c$ compared to the pristine Bi-2223 pellet. Our study suggests that larger sized agglomeration of magnetic nanoparticles of Co$_2$C leads to loss of superconductivity in the nanocomposite. However, there are also unagglomerated Co$_2$C nanoparticles distributed uniformly throughout the nanocomposite which acts as efficient pinning centres allowing for collective vortex pinning centres to be retained, even upto temperatures near $T_c$, and these nanoparticles also do not compromise the bulk $T_c$ of the superconducting fraction. Our study shows that these nanocomposites which exhibit enhanced $J_c$ especially in the high T regime are potentially useful for high current applications.

**Introduction:**

The critical current density ($J_c$) of type II superconductor, is the maximum dissipationless current density a superconductor can sustain. It is a crucial quantity for applications since it clearly indicates how strongly vortices are pinned in a superconductor. Pinning localizes the vortices and prevents dissipation from appearing in a superconductor until $J \leq J_c$ [1,2]. In realistic superconductors, the vortices are trapped or pinned on these sites which can be regions with local crystalline imperfections like regions with dislocations, vacancy sites or sites with impurity atoms[3,4]. In a superconductor, impurities and/or defects are either present naturally or are artificially introduced in the material[1,4,5,6]. It is well known that strong pinning centers are produced via heavy ion irradiation of the superconducting samples and also by artificial nano- structured patterning of superconductors[7]. Often nano - pinning centers whose size is comparable to ξ (the superconducting coherence length) or λ (the superconducting penetration depth), produce strong interaction between the pinning center and flux lines resulting in strong pinning and an enhanced $J_c$[8,9]. There has always been interest in exploring ways to increase the $J_c$ of HTSC (high transition temperature ($T_c$) superconductor) materials, as well as the recent Iron Pnictides superconductors[10,11,12,13,14]. Typically, atomic defects, oxygen vacancies, edge defects, impurities, grain boundaries, produce local variation in superconducting order parameter on the scale of 1-2 nm ~ ξ in HTSC, thereby producing strong point pinning centers [4-6,15,16]. Research groups have found an enhancement in $J_c$ by a factor of 2 to 7, by mixing various magnetic nanoparticles like $Co_3O_4$, $Fe_3O_4$, $BaZrO_3$ of different size in HTSC's [8,17,18]. However, the efficiency of magnetic nanoparticles (NP's) as pair breakers and hence acting as strong pinning centers often depends on the pairing symmetry as well as the gap structure of the superconductors[19,20,21]. The size, type, and concentration of the magnetic nanoparticles admixed into the superconductor also determine the extent to which the pinning force will increase[22]. Amongst high $T_c$ superconducting materials, $Bi_2Sr_2Ca_1Cu_2O_8$ (Bi-2212) and $Bi_2Sr_2Ca_2Cu_3O_{10}$ (Bi-2223) superconductor, the latter has a higher $T_c$ of 110 K[23]. Despite the challenges of making these materials into wire or tape form[24,25], they are important for their potential application in power transmission cables and high current carrying leads which operate at liquid Nitrogen temperatures[25,26]. As Bi-2223 has higher $T_c$ of 110 K, hence in recent times efforts are on to better explore the nature of its superconducting state in Bi-2223 with inequivalent $CuO_2$ planes as well as explore ways to enhance its $J_c$ by doping it [23,27,28,29,33].

Nanomaterials of transition metal carbide systems (TMCs, M= Fe, Ni, and Co) are well-known materials with wide applications in ferrofluids, catalysis, bioimaging, and memory devices[30,31,32]. The effects of combining TMCs with HTSC materials on $T_c$ and the superconductor's pinning properties have not been well investigated. Cobalt carbide nanoparticles have gained popularity recently among TMCs due to their high coercivity and remanence at ambient temperature. They appear in two forms, $Co_3C$ and $Co_2C$. While $Co_3C$ is known to be a ferromagnet, bulk $Co_2C$ is expected to be a paramagnet[30,33]. While in bulk form, $Co_2C$ is expected to be a paramagnet, some previous studies have suggested that nanoparticles of $Co_2C$ are ferromagnetic[34,35]. Recent studies however reveal[36,37] rich and complex magnetic properties of the 40 nm $Co_2C$ nanoparticles (NP's). The $Co_2C$ nanoparticles (40 nm) are ferromagnetic (FM) with strong magnetic anisotropy and with a high blocking temperature ~ 463 K. Along with ferromagnetism the $Co_2C$ nanoparticles exhibit the novel Exchange bias effect[38,39] and the presence of a low-temperature spin glass state[37]. The complex magnetic behaviour arises from a FM core and a disordered spin configuration in a thin shell of the $Co_2C$ nanoparticles. Given the complex magnetic nature of $Co_2C$ nanoparticles we were interested in exploring their effect on the pinning properties of HTSC's. In this paper we study the superconducting properties of NP composite pellets made using 15 nm diameter pristine Bi-2223 powder ($T_c$ = 109 K) mixed with 0.05% wt and 2% wt $Co_2C$ nanoparticles (40 nm diameter) powder. The ferromagnetic-superconductor (FM-SC) composite pellets were prepared by subjecting the powders to mechanical pressure without any additional heat or chemical treatment. A comparison of the magnetization (*M*) versus temperature (*T*) and *M* versus field (*H*) measurements of pristine and composite pellets, shows all the pellets possess similar $T_c \sim 109$ K. The magnetization response of these composite samples is complex, where the FM contribution to magnetization *i.e.,* $M_{FM}$ and the superconducting contribution, i.e., $M_{SC}$ are mixed. We analyse *M*(*T*) and *M*(*H*) data to separate out the two contributions. Our study shows about 97% ,97.3% and 98.34% loss by weight of superconducting fraction in the nanocomposite pellets with increasing admixture of $Co_2C$ for 0.05%, 2% and 10% respectively. Our analysis as a function of *H* and *T* shows that although there has been a significant loss of the superconducting fraction, the shape of the extracted behavior of the magnetization of the superconducting ($M_{SC}$) phase is found to be identical to that of the original superconducting Bi-2223 pellet. This indicates that the FM-SC composite retains the bulk superconducting pinning properties. We see that the superconducting phase in the composite exhibits an enhanced irreversibility, especially in the high *T* regime near $T_c$. At 77 K compared to the pristine Bi-2223 pellet, the irreversibility field increases by almost three times in the composite pellet. From our analysis of the $J_c(H,T)$ extracted from the $M_{SC}(H,T)$, the superconducting phase shows the presence of a strong pinning single vortex pinning (SVP) regime which crosses over at a field $H^*$ into a field dependent collective vortex pinning (CVP) regime. Notably, we see a significant enhancement in $H^*$ value for the composite pellet in the high *T* regime of 77 K, compared to the pristine Bi-2223 pellet where the $H^*$value is negligibly small at a similar *T* range. We find in the composite pellet at 77 K the appearance of a strong single vortex pinning regime extending from $0 < H \leq H^*$, while this regime is nearly absent in the pristine Bi-2223 pellet. A comparison of the composite with the pristine Bi-2223 pellet in the low *H* regime, shows the $J_c$ of the composite increases by a factor of almost 10 times at low *T* and we observe an increase by almost 100 times at high *T* (77 K). Our studies show while the superconducting phase in the SC - FM composites made with Bi-2223 NP's although have a reduced superconducting volume fraction, their pinning properties get significantly enhanced, especially in the high *T* regime near $T_c$. Such composites are potentially useful for developing superconductors for high current applications.

**$Co_2C$ and Bi-2223 nanoparticles and their nanocomposite preparation:**

Cobalt acetate tetrahydrate [$Co(CH_3CO_2)_2 \cdot 4\ H_2O$], tetra ethylene glycol (TEG), sodium hydroxide (NaOH), polyvinylpyrrolidone (PVP), and absolute ethanol from Sigma-Aldrich Co. are used as primary chemicals to synthesize the cobalt carbide nanoparticle clusters. We synthesize the samples via the one-pot polyol reduction process[36,37]. At first, 0.8 g of sodium hydroxide (NaOH) is dissolved into 20 ml TEG in a glass beaker (Borosil) by heating it to 100 °C. Another 30 ml of TEG is poured into a 100 ml European flask containing 2.5 mmol of $Co(CH_3CO_2)_2 \cdot 4\ H_2O$ and 0.75 g of PVP. The whole mixture is then stirred for 20 minutes at room temperature by using a magnetic stirrer. After mixing it for 20 minutes, it becomes a homogeneous mixture, and then NaOH is poured into the European flask, and we heat the mixture to 373 K for 30 min to remove water from the solution. The solution is heated further till it reaches its boiling point of TEG (583 K) for 1 hour. Then we allow the solution to cool down to room temperature naturally. After that, the solution is poured into methanol, in a test tube and the nanoparticles are precipitated. The precipitated nanoparticles are separated from the solution using a rare earth magnet. The residual solution is drained, and ethanol is added to the precipitate, and this process is repeated for several times. Later, using a vacuum oven, the extracted nanoparticles are dried, and the nanoparticle powders are compacted into pellets weighing 4 mg[36,37].

Polycrystalline pieces of Bi-2223 (CAN superconductors, Czech Republic) are ground into a fine powder using alumina pestle mortar continuously for 5 hours and repeated for 3 cycles (microstructure details of the polycrystal from which the powders are ground are presented later in fig.2). We prepared two batches using the powders of Bi-2223 and $Co_2C$ NP's: Batch-1 has Bi-2223 powder mixed with 0.05% by weight and batch-2 has 2% by weight of $Co_2C$ powder, respectively. The powders were mixed continuously for 4 to 5 hours in a pestle mortar and the process was repeated 3 times. A portion of the mixture from each batch was pressed with a hydraulic press to make pellets with dimensions 3 mm × 2 mm × 0.5 mm (batch-1, pellet with 0.05% by wt. of $Co_2C$, 14.4 mg) which we will refer to as 0.05%-Bi-2223 and 4.5 mm × 3 mm × 0.5 mm (batch 2, pellet with 2% by wt. $Co_2C$, 31.7 mg) which we will refer to as 2%-Bi-2223, also give information about the 10% pellet which we refer to as 10%-Bi-2223. $Co_2C$ dissociation temperature is between 500 to 600°C, so the composite pellets weren't subjected to high-temperature (beyond 450 °C) sintering. We also used for our measurements pristine $Co_2C$ pellet with dimensions of 2.13 mm × 1.23 mm × 0.36 mm (4 mg) and pristine Bi-2223 (0% $Co_2C$) pellet with dimensions of 2.8 mm × 2.2 mm × 1 mm (25.5 mg), which we will refer to as 0%-Bi-2223. Note that the 0%-Bi-2223 pellet was also prepared from the same batch of pristine Bi-2223 ground powder.

**Experimental Details and Microstructure investigation of the pellets:**

For preliminary characterization of the samples, Powder X-ray diffraction (XRD) was performed on the as-synthesized powder of $Co_2C$ nanoparticles using a Panalytical X-ray diffractometer with Cu $K_α$ radiation (λ=1.5406 Å) at room temperature in the 2-theta range of (30° − 90°) as shown in Fig 1(a). The same has been done for Bi-2223 pristine powder for XRD characterization in the range of (5° − 60°) as shown in Fig: 2(a). Electron Dispersive X-ray (EDX) measurements, using Instrument Model: JSM-6010LA; JEOL, were performed to get the distribution of basic elements (Bi, Sr, Pb, Ca, Cu, O and Co, C) of the admixture samples. For magnetic measurements of the samples, we used the SQUID magnetometer facility of IIT-Kanpur from Cryogenics, UK. Transmission Electron Microscopy (TEM) of the $Co_2C$ nanoparticles was performed using Model: FEI-Titan G2 60-300 KV TEM, from Advanced Imaging Centre, IIT-Kanpur. EBSD (Electron Backscattered Diffraction) measurements were performed to image the grain size of the Bi-2223 polycrystal samples using the EBSD facility available at ACMS, IIT-Kanpur, model: JSM-7100F; JEOL.

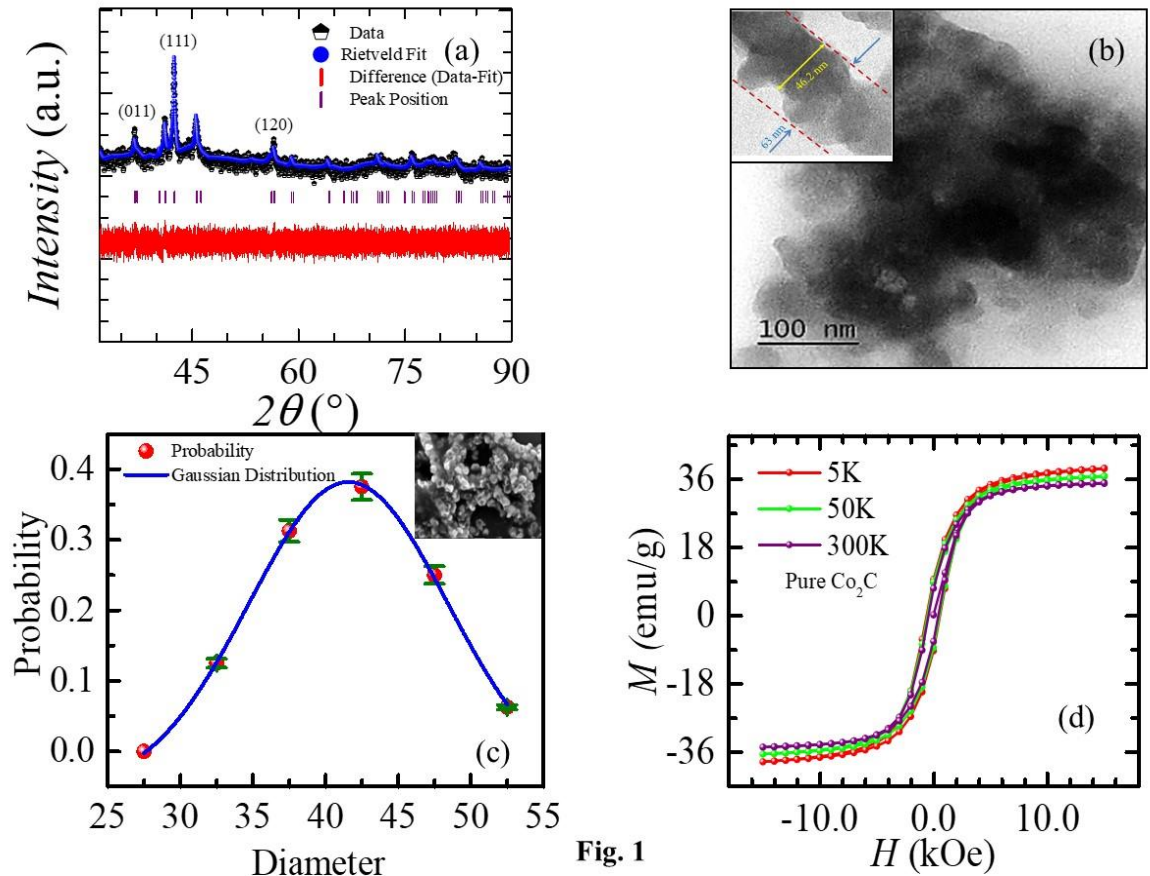

Fig. 1: (a) Peak position at XRD intensity data shows the signature of $Co_2C$ nanoparticles. (b) TEM images of $Co_2C$ nanoparticles. (c) Histogram of size distribution of $Co_2C$ nanoparticles. (c, inset) SEM images of clusters of $Co_2C$ nanoparticles. (d) Magnetic Hysteresis loop of $Co_2C$ nanoparticles for 5 K, 50K, and 300 K. It shows Ferromagnetic nature upto 5K.

Following the procedure described, we synthesize powders of Co$_2$C nanoparticles with an average diameter of approximately 40 nm (Fig 1(b)). The powder XRD pattern fits well with Rietveld refinement (using Fullprof package) and the fit corresponds to pristine Co$_2$C phase in the powder. The lattice parameters obtained for our powder are $a = 4.3797 \pm 0.0025$Å, $b = 4.4632 \pm 0.0025$Å and $c = 2.9030 \pm 0.0025$Å, which are within 0.3% of the standard values for Co$_2$C. The average diameter of the Co$_2$C nanoparticles (NPs) in the powder is determined using the Debye Scherrer equation: $D = \frac{\xi \lambda}{\beta \cos \theta}$, where $D$ is the average crystallite size, $\xi$ is the dimensionless shape factor ~ 0.9, $\beta$ is the full width at half maximum of XRD peak, and $\theta$ is the Bragg angle. Using the value of $2\theta = 42.5°$ and $\beta = 0.28°$, we estimate $D = 30.7$ nm $\pm$ 7 nm. The TEM image (Fig.1(b)) of the synthesized nanoparticle and a histogram analysis (Fig.1(c)) of the particle sizes, suggest an average size in the range of 40 $\pm$15 nm. The TEM image in Fig.1(b) and the SEM image in Fig.1(c, inset) show the tendency of the magnetic Co$_2$C nanoparticles to agglomerate into clusters. The clustering leads to the collective magnetic behaviour and strong magnetic moment associated with each cluster of Co$_2$C. The magnetic features of these nanoparticles have been explored in detail earlier[37]. The indexing of peaks seen in the powder XRD pattern fits well to the Bi-2223 phase (see fig.2(a)) [40] and (see supplementary section I) chemically confirms a Bi$_2$Sr$_2$Ca$_2$Cu$_3$O$_8$ (Bi-2223) stoichiometry of the powder, with a small concentration of Pb. From the width of the XRD peak at 59.63° in Fig. 2(a), the average size of crystallites in the Bi-2223 powder is ~ 15.3 $\pm$ 0.1 nm. Inset of Fig. 2(a) shows an SEM image of a region on the surface of 0%-Bi-2223 pellet. The original polycrystalline pellet was also characterized prior to grinding it to form the powder. EBSD image (fig.2(b)) of this polycrystalline pellet of the 0%-Bi-2223 shows grain of size in the range of 2-10 $\mu m$ (Fig. 2(b)) and they are randomly oriented having different crystal orientations (Fig. 2(c)). Analysis of the grain size distribution in fig. 2(d) showed log-normal distribution with a mean value~(5.18$\pm$0.01)$\mu$m. EDX mapping of the spatial distribution of the Co and C (see Figs. 2(e), (f) and (g), (h)) in the composite pellets showed a nearly uniform distribution, with low density of clustering of Co (larger sized pink spots) due to the agglomeration of Co$_2$C NP's. It may be noted from figs.1(e) and 1(f), that while we see blobs of Co suggesting presence of aggregates of Co$_2$C NP's, there is also a uniform distribution of pink speckles of Co distributed across the scan, suggesting also a uniform dispersion of Co$_2$C NP's across the pellet. The blobs we believe arises from the Co$_2$C strong tendency to cluster. Between 0.05%-Bi-2223 and 2%-Bi-2223 pellet there is a 40 times increase in the admixed amount of Co$_2$C (by weight). Analysis of a number of EDX elemental scan analysis similar to those in figs. 2(e) and (f) reveals that the average number density of the Co$_2$C cluster's changes from ~ 0.022 clusters/$\mu m^2$ in 0.05% pellet to ~ 0.025 clusters/$\mu m^2$ in the 2% pellet, namely, an increase in the number density of large sized clusters (blobs) by only ~ 10%, in going from 0.05%-Bi-2223 to 2%-Bi-2223 pellet. The remaining Co$_2$C- NP's are uniformly dispersed in the Bi-2223 NP composite.

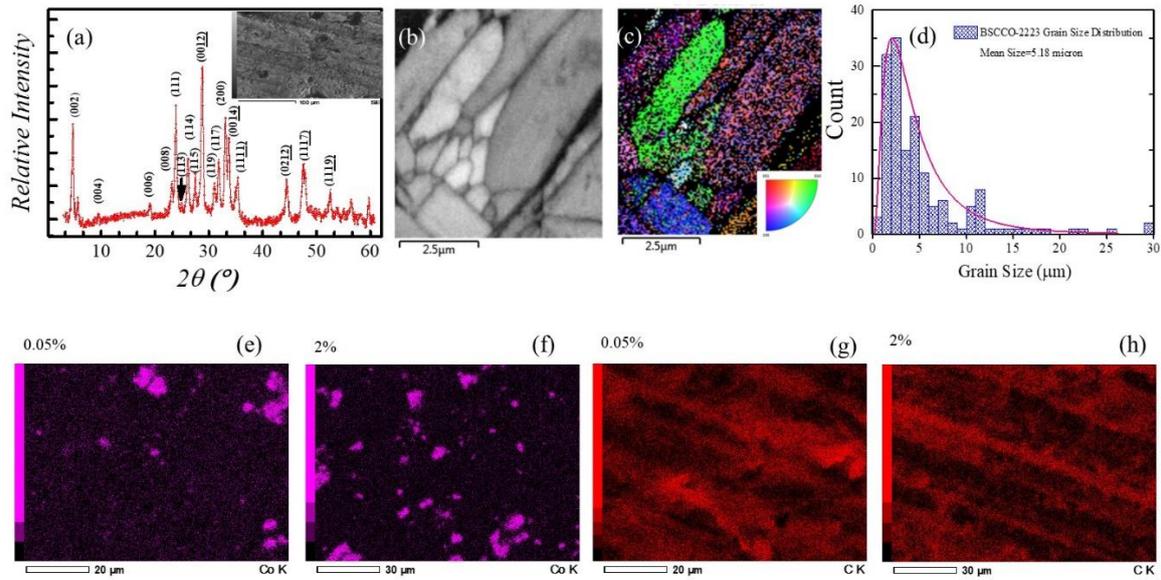

Fig. 2

Fig. 2: (a), Peak Positions of XRD intensity plots of Bi-2223 pellet after grinding the powders for 15 hrs (Inset shows SEM image of Bi-2223) .(b) EBSD image of the original polycrystalline Bi-2223 pellet (which was ground to obtain the Bi-2223 nano-powder). It shows the grains of the polycrystal (c) EBSD image shows the crystal grain orientation image. It shows each grains are oriented randomly, confirming its polycrystallinity of the original samples, (d) Distribution of the grain size within the original pristine Bi-2223 polycrystalline sample, and we see its grain size distribution follows a log-normal distribution, with a mean grain size of 5.18 micron ,(e) EDS image of Cobalt (Co) in Bi-2223 admixed with 0.05% $Co_2C$ and (f) ) EDS image of Cobalt (Co) in Bi-2223 admixed with 2% $Co_2C$ respectively, showing homogeneous distribution of Co- across the samples (pink speckles in the images), with some larger sized pink blobs (larger sized clustering of $Co_2C$ NP's).(g) EDS image of Carbon (C) in Bi-2223 admixed with 0.05% $Co_2C$ nanoparticles (h) EDS image of Carbon (C) in Bi-2223 admixed with 2% $Co_2C$ nanoparticles .

**Magnetization measurements of the nanocomposite pellets:**

The magnetization (*M*) versus applied field (*H*) (*M-H*) hysteresis loops measured at different temperatures (*T*) for the $Co_2C$ NP pellet is shown in Figure 1(d). The loops show FM in the NPs surviving up to room temperature (300 K). The FM hysteresis loop widths remain nearly constant from 5 K up to 300 K. We measure the magnetization of the pellets containing 0%, 0.05%, and 2% Bi-2223. For low-field *M-T* measurements, the pellets were first zero field cooled down to 2.3 K and then a low field of 100 G was applied. From the *M(T)* behaviour in fig.3(a) (and inset of fig.3(b)) for 0%-Bi-2223, 0.05%- Bi-2223 and 2%- Bi-2223 pellets, note that with the addition of $Co_2C$ the superconducting transition temperature $T_c$ of the pellets remains unchanged. From the onset of diamagnetism in *M(T)* we estimate $T_c$ ~ 109 ± 0.5 K. Figure 3(b) inset shows the *M(T)* behaviour near $T_c$ for the 0.05% and 2% pellets normalized by their respective mass, i.e., e.m.u/gm units, and in the main panel of fig.3(b) we show the *M(T)* data in un-normalized (e.m.u units). Figure 3(b) shows the expected saturated Meissner diamagnetic *M* signal at low *T*. The *M* sharply increases from this saturated value with increasing *T* near $T_c$. A new feature in fig.3(b) main panel which is not seen in the typical behaviour of *M(T)* of pristine superconducting systems, is the $M(T^*) \sim 0$ at $T = T^*$ which is less than $T_c$, and at $T > T^*$ the *M(T)* increases and saturates to a finite positive *M* above $T_c = 109$ K. This feature seen between $T^*$ and $T_c$, is due to the FM contribution to *M* from the $Co_2C$ fraction competing with the *M* contribution from the superconducting fraction present in the composite pellet. In the normal state above $T_c = 109$ K the FM, positive *M* response from $Co_2C$ dominates, so that the net *M(T)* response is positive (see Fig.3(b) main panel). In fig.3(b) the un-normalized e.m.u scale for the main panel helps to show that the positive magnetization signal changes in proportion to the amount of $Co_2C$ (0.05% or 2%) admixed in the composite pellet. As *T* decreases below $T_c = 109$ K, the diamagnetic *M* contribution to the net measured *M* signal from the superconducting fraction increases until it completely compensates the positive *M* contribution at $T^*$. Figure 3(b) main panel shows that at *T** of 99.5 ± 0.2 K (for 0.05%-Bi-2223 pellet) and 93.3 ± 0.2 K (for 2%-Bi-2223 pellet) the two responses completely balance resulting in $M(T^*) \to 0$. To separate out the contributions to the net *M* from the FM component (i.e., $M_{FM}$) and the superconducting contribution, i.e., $M_{SC}$, to a first approximation, we consider $M = M_{SC} + M_{FM}$. Since we already know from fig.1(d) that $M_{FM}$ doesn't change significantly with *T*. At $T > T_c = 109$ K, we consider the $M = M_{FM}$ as $M_{SC} = 0$ here. Figure 1(d) shows that $M_{FM}$ of $Co_2C$ NP's is only weakly *T* dependent (the FM loops change very slightly between low *T* and 100 K), therefore in fig. 2(c) we determine that $M_{FM}(T < T_c)$ ~ 3 x $10^{-4}$ emu (for 0.05%-Bi-2223) and $M_{FM}(T < T_c)$ ~ 9.5 x $10^{-4}$ emu (for 2%-Bi-2223). Using these values in fig.2(c) we plot $M_{SC}(T) = (M(T) - M_{FM})$ for both composite ($Co_2C$ - Bi2223) pellets and compare it with that for 0%-Bi-2223 pellet. In fig.3(c), we see that the saturated Meissner *M(T)* signal at $T = 2.3$ K for the 0.05%-Bi-2223 and 2%-Bi-2223 pellet is significantly suppressed compared to the pristine 0%-Bi-2223 pellet. The suppression is directly related to a loss of superconducting fraction due to the presence of FM pair breakers in the SC-FM composite pellets.

To estimate the change in the superconducting volume fraction in the SC-FM pellets, we use the $M_{SC}(T)$ values to calculate the ratio $\hbar(T) = \frac{M_{sc}(T, 0\%-Bi-2223)}{[M_{sc}(T, 0.05\%-Bi-2223) \text{ or } M_{sc}(T, 2\%-Bi-2223)]}$. Parameter $\hbar(T)$ measures the ratio of the superconducting volume present in the pristine pellet to that in the composite pellet. From fig. 3(d) we see that at 2.3 K the Meissner superconducting volume of pristine 0%-Bi-2223 pellet is larger than that in of 0.05%-Bi-2223 pellet by almost 33 times and 2%-Bi-2223 pellet by 38 times, respectively. Thus, the ferromagnetism of $Co_2C$ causes a loss of superconducting volume fraction of 97% in the 0.05%-Bi-2223 pellet and a loss of about 97.3 % in the 2 %-Bi-2223 pellet when compared to the original pristine 0%-Bi-2223 pellet. We have also studied the $M_{sc}$ vs $T$ of a 10%-Bi-2223 pellet, viz., with 10% $Co_2C$ NP's powder admixed by weight in the pristine Bi -2223 powder (see the comparison of $M_{sc}$ vs $T$ for 0%, 0.05%, 2% and 10%-Bi-2223 pellets in the supplementary section 1). Based on similar analysis as above, we find that mixing 10% by wt. of $Co_2C$ NP's powder in Bi-2223 powder, the superconducting volume fraction gets suppressed by 98.34% i.e., only 1.66% superconducting volume fraction survives in the sample. It must be mentioned that although there is a loss of superconducting volume fraction, the fraction which retains superconductivity exhibits robust Meissner effect and diamagnetism signatures characteristic of bulk superconductivity. Subsequently we will investigate the pinning in the surviving superconducting phase. Between 0.05% and 2% pellet we observe a 10 % additional loss of superconducting volume fraction although there is a 40 times increase in the weight percentage of $Co_2C$ in the nanocomposite pellet. Clearly the additional loss of superconducting volume fraction has only a weak dependence on the weight percentage increase of $Co_2C$ in the admixed pellet. This is because as already discussed earlier, the weight percent change of $Co_2C$ in Bi - 2223 translates to a slow increase in the density $Co_2C$ larger sized clusters (blobs). Our earlier analysis of the EDX data revealed a change in number density of clusters of about ~ 10% which agrees well with the 10% drop in the superconducting fraction in going from the 0.05%-Bi-2223 to 2%-Bi-2223 pellet. Thus the agglomerated clusters (blobs) of $Co_2C$ in the nanocomposite are responsible for destroying bulk superconductivity. These large blobs of $Co_2C$ are strong ferromagnetic macroscopic regions in the superconducting medium which destroy superconductivity around them. However, the finely dispersed nanoparticles of $Co_2C$ amongst the Bi-2223 (fine pink speckles seen in figs. 2(e) and 2(f)) are likely to influence the pinning properties without destroying superconductivity, a feature we explore in greater detail subsequently. We find that $\hbar(T)$ decreases with increasing $T$. Note that the behaviour of superconducting $\hbar(T)$ with increasing $T$ mimics the $M(T)$ behaviour of the pristine Bi-2223 superconductor. The rate at which $\hbar(T)$ drops is determined by ratio of the rates at which $M_{SC}(T)$ for the pristine and composite pellets decreases towards zero with increasing $T$ near $T_c$. The $\hbar(T)$ is identical for both 0.05%-Bi-2223 and 2%-Bi-2223 pellet, and both curves extrapolate down to zero at a $T \sim 110$ K $\approx T_c$. It is clear that by mixing the $Co_2C$ NP's in Bi-2223, although there is some loss of superconducting phase due to clustering of $Co_2C$, the $T_c$ of the retained superconducting fraction is however unaffected compared to the pristine 0%-Bi-2223 pellet. We now investigate the effect on pinning in the superconducting fraction due to the presence of $Co_2C$ NP's distributed across the pellet.

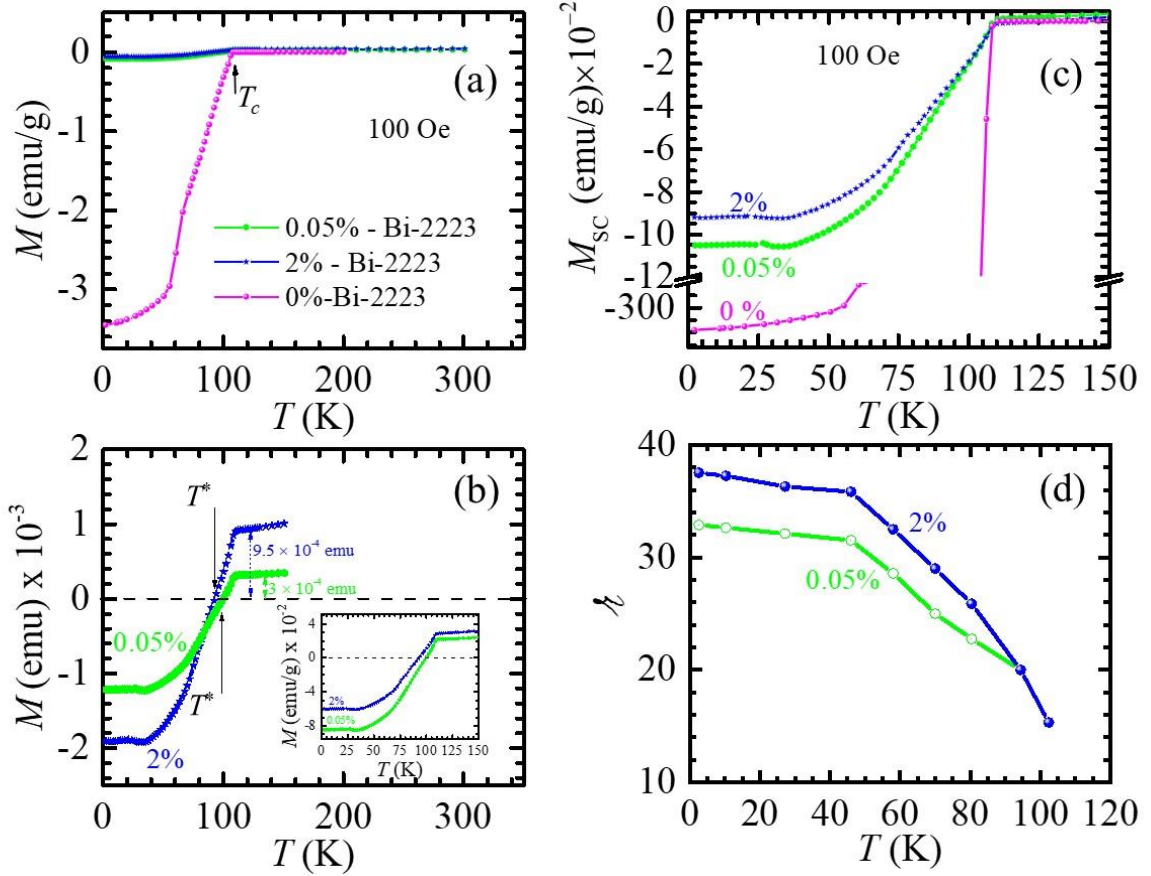

Fig 3: (a) Magnetization as function of temperature at 100 Oe of pure Bi-2223, 2% $Co_2C$ added with Bi-2223 (2%-Bi-2223) and 0.05% $Co_2C$ added Bi-2223 (0.05%-Bi-2223) where $T_c$ shows the transition temperature. (b) Zoomed view of Magnetization (in emu unit) as function of temperature of 2%-Bi-2223 and 0.05%-Bi-2223. $T^*$ is the temperature where magnetization value becomes zero. The values of offset magnetization are $9.5 \times 10^{-4}$ emu and $3 \times 10^{-4}$ emu for 2%-Bi-2223 and 0.05%-Bi-2223 respectively. (b (inset)) Shows the $M(T)$, normalized by its mass for the 0.05%-Bi-2223 and 2%-Bi-2223 pellets in an expanded scale respectively. (c) Magnetization after subtraction of offset magnetization as function of temperature compares them with the $M(T)$ for 0%-Bi-2223 pellet (shown with a y-axis break). (d) A ratio parameter of magnetization of the pristine superconducting sample to the response from the superconducting part of the nanocomposite (see text for details), namely, $\lambda$ $= M_{SC}(0\%)/M_{SC}$(mixture), plotted as a function of temperature($T$) for 2%-Bi-2223 (blue) and 0.05%-Bi-2223 (green) pellets.

**Exploring the pinning properties of the nanocomposite and pristine pellets:**

In order to explore the pinning properties of the nanocomposite and pristine Bi-2223 pellets, we analyse the magnetization hysteresis loops of these systems. The Figs. 4(a) and (d) shows the five quadrant $M(H)$ hysteresis loop for the 0.05%-Bi-2223 pellet at 5 K and 77 K, respectively. The low field portion of the virgin forward $M(H)$ leg of the curve (shown with a blue arrow marked in Fig. 4(a) shows a linear $M(H)$ behaviour. The zoomed-in portion of the virgin linear $M(H)$ is shown in the inset of Fig. 4(a). Note that above a field marked the penetration field, $H_p$, the $M(H)$ curve begins to deviate from linearity and the diamagnetism decreases due to the penetration of vortices in the pellet. The deviation from linear $M(H)$ behaviour represents the presence of a penetration field $H_p = 0.067$ T (at 5 K), which is typical feature of bulk superconductors. Usually $H_p$ is higher than $H_{c1}$ due to demagnetization, surface barrier, uniformity of sample edges, effects[41]. The presence of a penetration field feature along with a linear $M(H)$ curve at $H < H_p$, suggests the presence of robust Meissner diamagnetic response in the composite (similar feature is also seen in fig.5 for the 2%-Bi-2223 pellet). The presence of a robust Meissner effect feature in $M(H)$, along with signatures of a bulk penetration field suggests that although in the SC-FM composite pellet, the superconducting volume fraction has shrunk, the retained superconducting phase in the pellet still possesses features of robust bulk superconductivity with a well-defined bulk $T_c$. The nature of the superconducting response we observe is one of macroscopic phase coherent superconductivity and is not associated with features related to weak superconducting fluctuations present locally in the pellet and where macroscopic phase coherent superconductivity is absent. The $M(T)$ of the 0.05%-Bi-2223 pellet shows that instead of $M(H)$ remaining in the fourth quadrant of the plot (viz., in the diamagnetic sector with a -|M| value), it moves into the first quadrant (+|M| value) for $H > 0.5$ T (see inset and main panel of Fig. 4(a)). While the curve displays a hysteresis between the forward (*f*) and reverse (*r*) legs, the overall shape of the $M(H)$ loop is distorted compared to the *M-H* curve of a pristine HTSC superconductor. Here too the observed distortion of the $M(H)$ curve is related to the FM $Co_2C$ NP's contributing to $M(H)$. In order to extract out the magnetization versus *H* response of the superconducting fraction, i.e., $M_{SC}(H)$, we use a method which has been employed in the past for other superconducting-magnetic systems[42]: From the *f* and *r* legs of the $M(H)$ hysteresis loop, we first determine the average *M*, i.e., $M_{avg}(H) = \frac{[M_f(H)+M_r(H)]}{2}$ for 5 K and 77 K, as shown in Fig. 4(b) and Fig. 4(e) (the data is normalized with the $M_{avg}$ value at ±6T). For comparison in fig.4(b) (red symbol) we overlay in the plot the $M_{avg}(H)$ of pristine $Co_2C$ NPs powder determine from the $M(H)$ fig.1(d). The close match seen between the two curves in fig.4(b) shows that the $M_{avg}(H)$ we have calculated for the composites (in Figs. 4(b) and 4(e)) is a reasonably good approximation of the behaviour of the average magnetization versus *H* behaviour of the pure FM $Co_2C$ NP's present in the pellet. The behaviour of $M_{SC}(H)$ is determined by subtracting the $M_{avg}(H)$ from the measured $M(H)$ hysteresis loop. The behaviour of $M_{SC}(H) = M(H) - M_{avg}(H)$ at 5 K and 77 K is shown in Fig. 4(c) and Fig. 4(f) for the composite pellet, 0.05%-Bi-2223. Note the shape of $M_{SC}(H)$ superconducting fraction in the composite pellet in figs. 4(c) and 4(f), is identical to the shape of the irreversible $M(H)$ hysteresis loops seen in our pristine 0%-Bi-2223 pellet measured at 5 K (see inset of fig.4(b)). This feature again confirms the macroscopic bulk superconductivity present in the superconducting fraction present in the composite pellet (and its not superconducting fluctuations). Due to the relatively weak hysteresis width of Co$_2$C NP's $M(H)$ about its $M_{avg}(H)$, it doesn't affect the $M_{SC}(H)$ behaviour we have obtained. It is important to note that Figures 4 and 5 represents the un-normalized M values in e.m.u units for the pristine 0%-Bi-2223 pellet as well as the 0.05 % and 2 % - Bi-2223 pellets (Fig. 5). Therefore, one should not attempt to compare the magnitude of the M values for the 0.05 % and 0 % -Bi-2223 pellets in these figures (and 0%-Bi-2223 pellet and 2%-Bi-2223 pellet in Fig. 5). Note that as the superconducting fraction is very small for the 10%-Bi-2223 pellet therefore the hysteresis in $M_{sc}(H)$ is small (especially at high $T$ in the range of 77 K). Hence we have restricted our analysis to the 0.05% and 2%-Bi-2223 pellet here, as well as for our subsequent analysis.

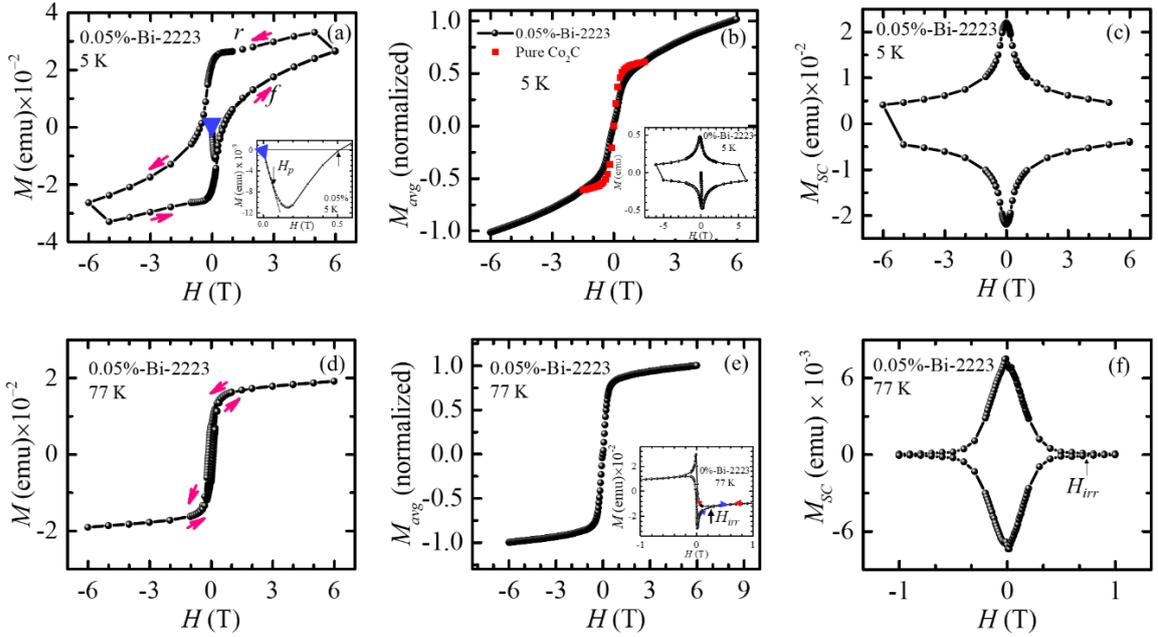

Fig. 4: All the Figures are for 0.05% of Co$_2$C added with Bi-2223 (0.05%-Bi-2223). (a) Five quadrants $M(H)$ loop at 5 K. Inset shows first quadrant of $M(H)$ loop and the deviation from linear $M(H)$ behaviour represents the penetration field $H_p$ of the superconductor (b) The black filled circles, show the behaviour of the calculated average magnetization as function of field at 5 K (see text for details) for the 0.05%-Bi-2223 composite. The magnetization data is normalized with the average M value determined at 6T. In the same plot. we have also shown with red symbols the average M value of pure Co$_2$C NPs powder determine from the $M(H)$ in fig.1(d). The Inset shows five quadrants magnetization loop of 0%-Bi-2223 at 5 K. (c) Magnetization as function of field after subtraction of average magnetization from forward run and reverse run of 5 K 0.05%-Bi-2223 magnetization loop. (d) Five quadrants $M(H)$ loop at 77 K. (e) Average magnetization of forward run (6 T to -6 T) and reverse run (-6 T to 6 T) for 77 K.

Inset shows five quadrants magnetization loop of 0%-Bi-2223 at 77 K. (f) Magnetization as function of field after subtraction of average magnetization from forward run and reverse run of 0.05%-Bi-2223 at 77 K.

Comparison of the $M_{SC}(H)$ at 77 K in fig.4(f) for 0.05%-Bi-2223 with that for the pristine 0%-Bi-2223 pellet in fig.4(e) inset, shows that by mixing of $Co_2C$ we observe a significant change in the irreversibility of the $M_{SC}(H)$ hysteresis loop at 77 K as compared to the pristine 0%-Bi-2223 pellet. In the inset of Fig. 4(e) we see that while the $M(H)$ hysteresis loop is nearly reversible at 77 K for the pristine 0%-Bi-2223 pellet, while there is significant irreversibility in the $M_{SC}(H)$ at 77 K for 0.05%-Bi-2223 (fig.4(f)). The width of hysteresis in $M(H)$ for the pure 0%-Bi-2223 superconductor at high $T$ (77 K) is significantly smaller than that at low $T$ (5 K) (see fig. 4(b)), as enhanced thermal fluctuations significantly blur the vortex pinning potential. In the inset Fig. 4(e) the red and blue arrows represent the overlapped virgin, forward and reverse legs of the $M(H)$ hysteresis loop above the irreversibility field of $H_{irr}$(77 K) = 0.3 T (see fig.4(e) inset). As the upper critical field is very high, therefore upto 1 T the pristine pellet exhibits a diamagnetic, reversible $M(H)$ response between 0.3T upto 1T. At 77 K below $H_{irr}$, the pristine pellet has a weak irreversibility (fig.4(e) inset). By comparing the shape of the $M_{SC}(H)$ hysteresis width for 0.05%-Bi-2223 pellet (Fig. 4(f)) with that of the 0%-Bi-2223 pellet (Fig. 4(e) inset), we see that at 77 K the vortex pinning strength of the superconducting state in the composite pellet at 77 K has significantly enhanced compared to the pristine pellet. Note that in 0.05%-Bi-2223 pellet the $H_{irr}$(77 K) has increased by three times to 0.9 T compared to that in the pristine 0%-Bi-2223 pellet. There is also a significant enhancement in the $M(H)$ hysteresis loop width at 77 K in the 0.05%-Bi-2223 pellet. Hence the effect of mixing $Co_2C$ in Bi-2223 on pinning in the high $T$ regime of the HTSC seems to be very significant, especially in the high $T$ regime. Note that similar analysis as above was for the 2%-Bi-2223 pellet (cf. Fig. 5). By determining $M_{avg}(H)$ from the $M(H)$ (following the same procedure outlined for fig.4), we obtain the behaviour of $M_{SC}(H)$ at different $T$ for the 2%-Bi-2223 pellet. The behaviour of $M_{SC}(H)$ at different $T$ seen in Fig. 5 for the 2%-Bi-2223 pellet is like that in Fig. 4 for the 0.05%-Bi-2223 pellet.

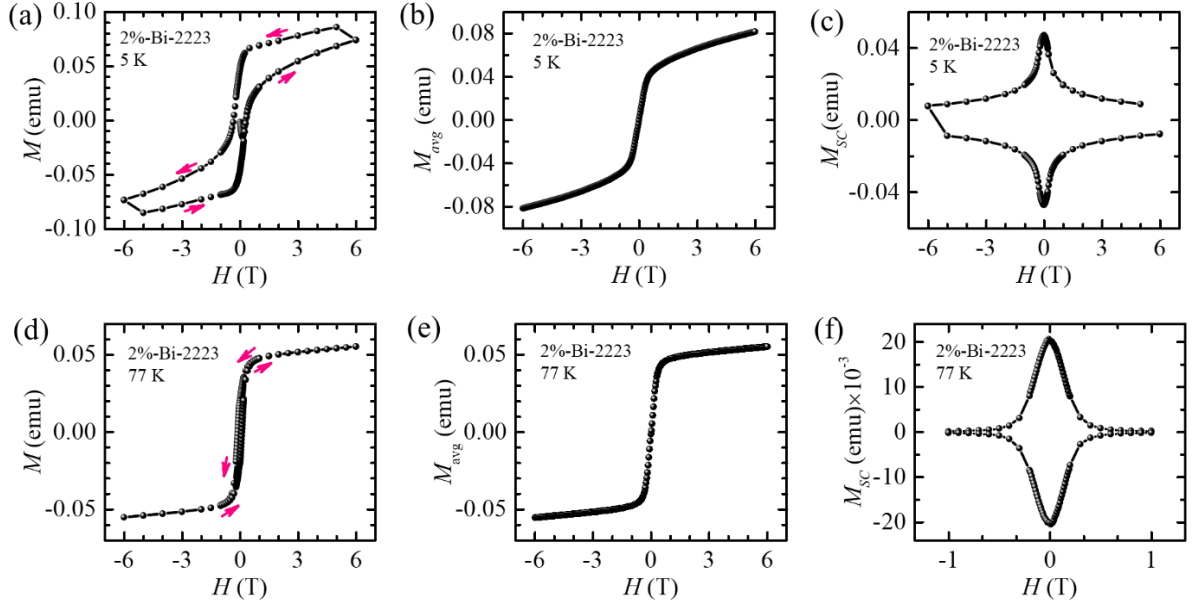

Fig. 5: All the Figures are for 2% of Co$_2$C added with Bi-2223 (2%-Bi-2223). (a) Five quadrants $M(H)$ loop at 5 K, (b) Average magnetization ($M_{avg}$) of forward run (6 T to -6 T) and reverse run (-6 T to 6 T) as function of field at 5 K, (c) Magnetization ($M_{SC}$) as function of field after subtraction of average magnetization ($M_{avg}$) from forward run and reverse run of 2%-Bi-2223 loop at 5 K. (d) Five quadrants $M(H)$ loop at 77 K, (e) Average magnetization of forward run (6 T to -6 T) and reverse run (-6 T to 6 T) as function of field at 77 K, (f) Magnetization ($M_{SC}$) as function of field after subtraction of average magnetization ($M_{avg}$) from forward run and reverse run of 2%-Bi-2223 at 77 K.

**Understanding the enhancement of $J_c$ in different pinning regimes of $Co_2C$– Bi-2223 nanocomposites:**

Using the $M_{SC}(H)$ determined for the composite as well as pristine pellets and superconducting volume fraction determined using the $\lambda(T)$ of Fig. 3(d), we estimate the width of the hysteresis, $\Delta M(H)$, in A.m$^{-1}$ units, at different $T$. From $\Delta M$ we estimate the bulk critical current density ($J_c$) as a function of $H$ at different $T$, using the expression $J_c = 20\,\Delta M/[a(1 - a/3b)]$ where $a$ and $b$ are the crystal dimensions perpendicular the applied $H$[43,44]. Using this for the 0%-Bi-2223 pellet the estimated $J_c = 10^6$ A.m$^{-2}$ (Fig. 6(a)) in nominally zero magnetic field at 77 K, compares well with the value of $J_c = 5.7 \times 10^6$ A.m$^{-2}$ at 77 K determined using $I$-$V$ measurements[45]. This shows that the process of mechanically grinding and pelletizing doesn't significantly modify the $J_c$ of the 0%-Bi-2223 pellet compared to the unground Bi-2223 polycrystal. It is immediately clear from fig.6 there is a significant enhancement in $J_c(H)$ over all $H$ and also at different $T$ (of 5 K and 77 K) for composite pellet. While we will return to this feature later; however below we analyse the $J_c(H)$ behaviour in greater details. For the 0%-Bi-2223 pellet at 5 K, the $J_c(H)$ shows a field-independent nature in the low $H$ regime (shown with the horizontal dashed line). This is the strong single vortex pinning (SVP) regime, where due to weak intervortex interactions at low $H$ (spacing between vortices $\gg \lambda$) the weakly interacting vortices get individually very strongly pinned[1]. In this regime, due to low inter-vortex interactions, the $J_c(H)$ exhibit weak H dependence. Increasing $H$ leads to stronger intervortex interaction between closely spaced vortices thereby enhancing the collective rigidity of the vortex state. This rigid vortex state exhibits weaker effective pinning. The CVP regime is characterized by a field dependence of the form, $J_c(H) \propto H^{-\beta}$, where $\beta$ is a positive constant[46,47,48,49,50,51,52]. We see in fig.6(a$_1$) that beyond a field $H^*$, there is a transformation from SVP regime to the collective vortex pining (CVP) regime. In the 0%-Bi-2223 pellet the value of $\beta \sim 0.45 \pm 0.02$ at 5 K is within the range of value $\beta$ found in the CVP regime[46,47,48,49,50,51,52]. At 77 K we find an enhanced $\beta \sim 1.49 \pm 0.08$ (~1.5) at 77 K. Note that at higher $T$ (as seen at 77 K) there is the added effect of thermally induced suppression of vortex pinning which produces a faster rate of suppression of $J_c$ with increasing $H$, i.e., $\left|\frac{d(\log J_c)}{d(\log H)}\right| = \beta$, compared to that at lower $T$. The effects of thermal smearing of pinning at high T is seen clearly in fig.6. For the pristine 0%-Bi-2223 pellet at 5 K as the effective pinning is much stronger, hence the SVP regime is clearly identifiable over a wide $H$ regime in fig.6(a$_1$) (solid green triangles in fig.6(a$_1$)) whereas at 77 K the SVP regime has shrunk and cannot be identified (Fig. 6(a$_2$) the solid pink coloured triangles). We see that in the $Co_2C$ NP's and Bi-2223 powder composite pellets, there is a significant enhancement in $J_c$ at both low (5 K) and high $T$ (77 K) (see figs. 6(a$_1$) and 6(a$_2$)). In fig.6(a$_1$) comparing the data for 0%-Bi-2223 in 0.05%-Bi-2223 pellet, shows only a slight increase in $H^*$ at 5 K, however the same comparison in fig.6(a$_2$) at 77 K shows a significant increase $H^*$ for the composite pellet. In fig.6(b) we compare the composite and pristine pellets at different $T$ the values of $J_c(H)$ by determining the ratio of $J_c$ for 0.05%-Bi-2223 ($J_{c,0.05\%}$) to that for 0%-Bi-2223 ($J_{c,0\%}$) pellet, i.e., $\frac{J_{c,0.05\%}}{J_{c,0\%}}$ (H). From fig.6(b) we see an increase in $J_c$ of the 0.05%-Bi-2223 pellet by almost 10 times compared to the pristine pellet and the increase in $J_c$ is uniformly retained at all H. At 5 K between the composite and pristine pellet, nature of the $J_c(H)$ curves are nearly similar with the $J_c(H)$ plot shifted for the composite. Hence the ratio $\frac{J_{c,0.05\%}}{J_{c,0\%}}$ (H) is feature less for 5 K.

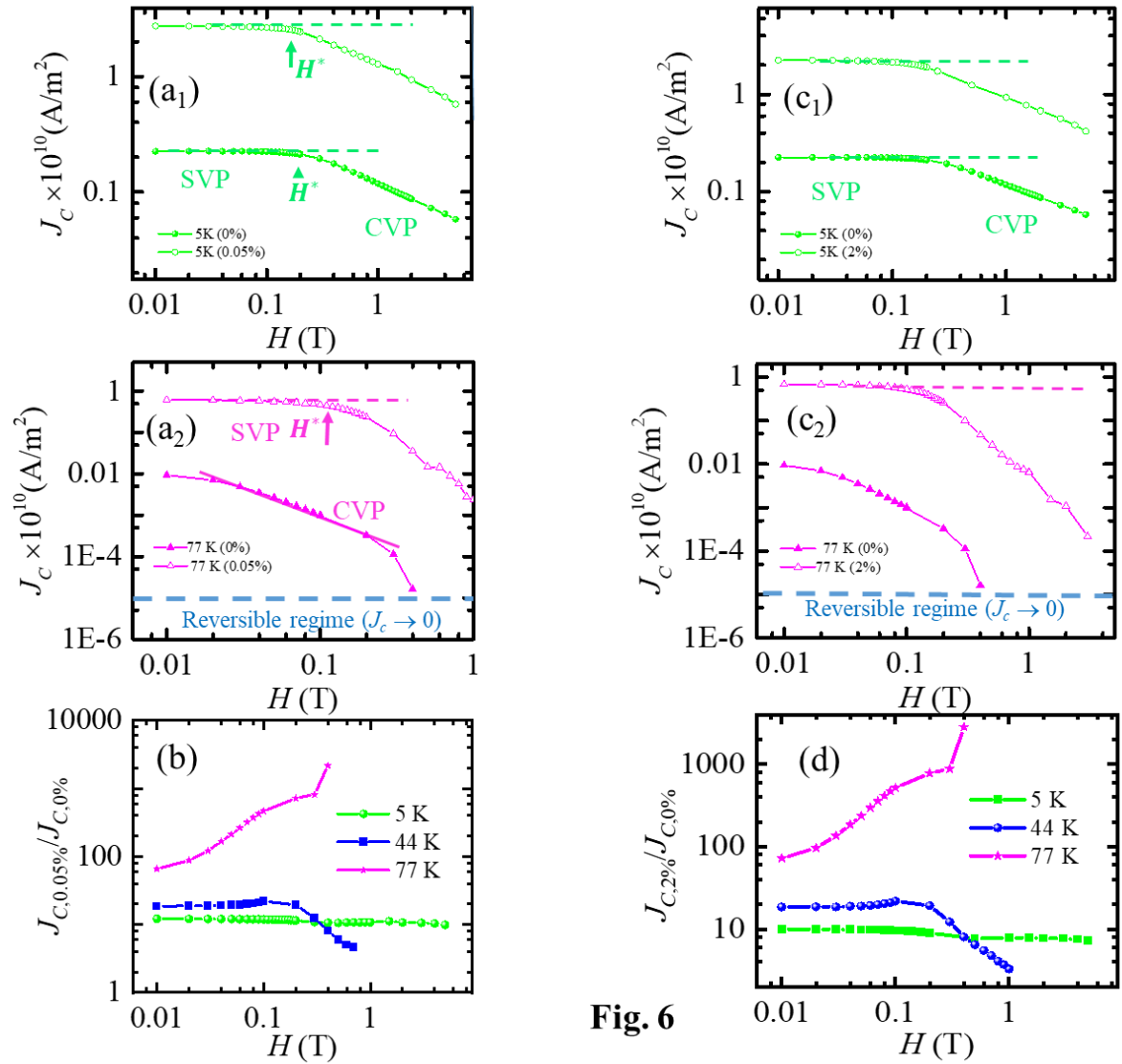

Fig. 6. ($a_1$) Critical current density as function of magnetic field at 5 K for 0.05% $Co_2C$ added with Bi-2223 (0.05%-Bi-2223) and for superconductor (0%-Bi-2223). ($a_2$) Critical current density as function of magnetic field at 77 K for 0.05% $Co_2C$ added with Bi-2223 (0.05%-Bi-2223) and for superconductor (0%-Bi-2223). (b) Ratio of critical current densities $J_{c,\ 0.05\%}/\ J_{c,0\%}$ at 5 K, 44 K, and 77 K. ($c_1$) Critical current density as function of magnetic field at 5 K for 2% $Co_2C$ added with Bi-2223 (2%-Bi-2223) and for superconductor (0%-Bi-2223). ($c_2$) Critical current density as function of magnetic field at 77 K for 2% $Co_2C$ added with Bi-2223 (2%-Bi-2223) and for superconductor (0%-Bi-2223). (d) Ratio of critical current densities $J_{c,\ 2\%}/\ J_{c,0\%}$ at 5 K, 44 K, and 77 K.

From fig.6($a_2$) we see that in the high T regime of 77 K there is a substantial increase in $J_c$ of the 0.05%-Bi-2223 pellet compared to the pristine 0%-Bi-2223 pellet. At 77 K the CVP regime in the 0.05%-Bi-2223 pellet extends upto a field of 1 T where the $J_c$ ~ 0.01 A.m$^{-2}$ while at 77 K in the pristine 0%-Bi-2223 pellet, by 0.4 T the CVP regime hits the reversible $J_c$ ~ 0. Thus, at high $T$ of 77 K in the 0.05%-Bi-2223 pellet the pinning strength is not smeared out by thermal fluctuations and has significantly strengthened. Furthermore, the field range over which pinning remains effective is significantly enhanced in the composite. Figure 6($a_2$) also shows that while in the pristine 0%-Bi-2223 pellet close to 0.3T, $J_c(H) \to 0$, in contrast the 0.05%-Bi-2223 the $J_c(H = 0.3T)$ remains finite upto ~ 1 T. This feature reconfirms what was seen earlier in fig.4(f), were at 77 K the composite pellet showed an increase in the irreversibility field by almost 3 times compared to the values in the pristine pellet. At 77 K in the composite pellet, the $\frac{J_{c,0.05\%}}{J_{c,0\%}}$ (H) plot in Fig. 6(b) shows an enhancement by a factor of 100 times at low H, which can reach upto almost 1000 times at higher $H$ (due to the expansion of the CVP regime in the composite samples). From fig.6(b) we see that in the SC-FM composite pellet (0.05%-Bi-2223) at 5K the enhancement in $J_c$ is by a factor of ten and is almost uniform at all $H$. However, the same composite pellet shows an order of magnitude greater enhancement in $J_c$ at higher T where the enhancement increases in high $H$ regime. In figs. 6($c_1$), 6($c_2$) and 6(d) we compare the $J_c$ behaviour for the 2%-Bi-2223 pellet and 0%-Bi-2223 pellet and here too we see similar features as that found in figs. 6($a_1$), 6($a_2$) and 6(b). In a table below, we have summarized all the important findings of both the batches of nanoparticle-superconductor composites. Recall here that the clustering of Co$_2$C NP's (pink blobs seen in figs.2(e) and 2(f)) in the nanocomposite is responsible for the loss of superconducting fraction (discussed earlier). The increase in $J_c$ of the retained superconducting fraction in the nanocomposite is a result of the Co$_2$C nanoparticles uniformly distributed in the matrix of Bi-2223 nanoparticles (the pink speckles in fig.2(e) and (f)). The ferromagnetism of these NP's is weak enough that they do not destroy superconductivity of the nanocomposite. However, the exchange interaction of these magnetic Co$_2$C NP's and their local fields are strong enough that they locally act as Cooper pair breakers for the superconducting condensate around their vicinity. Thereby, they locally suppress the superconducting order parameter of Bi-2223, without affecting the bulk $T_c$ of the superconducting fraction. It is this local order parameter suppression by the FM - NP's in the nanocomposite pellet which generates the strong pinning effects. The pins we believe are local point like features rather than extended pins, because in fig. 6 we see the clear evidence of collective vortex pinning regime, which is known to arise from collective interactions of vortices with point like pinning centre's[1,46]. We see that these Co$_2$C NP pinning effects are not smeared out due to thermal fluctuation effects which is a very significant effect in HTSC materials like Bi-2223. The effectiveness of the vortex pinning depends on, the ratio of the pinning potential well depth $U$ to thermal energy, and also the Ginzburg number ($G_i$)[1] for a superconductor ($G_i$ is the ratio of $T_c$ to the zero temperature superconducting condensation energy). The $G_i$ governs the width of the critical thermal fluctuation regime or the reversible regime in the field - temperature vortex phase diagram of the superconductors. Infact as the thermal smearing out of the pinning potential strength is significantly high in HTSC due to their large $T_c$ and the $G_i$ of HTSC is also high[1] (~ $10^{-1}$ to $10^{-2}$ for HTSC), all of these factors combine to weaken the pinning effects in most HTSC materials, especially in the high T regime (especially 77 K and above). However we find that the ferromagnetic $Co_2C$ - NP's act as efficient pinning centres, allowing for collective vortex pinning to survive even in the high temperatures regimes (both $H_{irr}$ and $J_c$ at 77 K in the nanocomposite is enhanced compared to pristine superconducting sample). One may recall that, raising the temperature causes thermal fluctuations to become more pronounced, thereby softening the vortex state; conversely, raising $H$ increases inter vortex interactions, which strengthen the vortex state's rigidity[1]. When $T$ increases and the effects of enhanced thermal fluctuations also haven't completely smeared out the pinning strength in the sample (i.e., pinning strength ≠ 0 at finite $T$), then thermally softened vortex lines find it easier to adapt to pinning centres and enter a strongly pinned vortex state regime. In contrast a more rigid vortex state is comparatively in a weaker pinned state[1]. From figs. 6(b) and 6(d) the behaviour of $\frac{J_{c,0.05\%}}{J_{c,0\%}}(H)$ or $\frac{J_{c,2\%}}{J_{c,0\%}}(H)$ at 44 K is intriguing. We observe that in the weak interaction regime of low $H$ (< 0.1 T) the thermally softened vortex state at 44 K accommodates the pinning centers of $Co_2C$ NP's well. As a result there is an increase in pinning as $T$ increases (from 5K → 44 K → 77 K) (see increase in $\frac{J_{c,0.05\%}}{J_{c,0\%}}(H)$, $\frac{J_{c,2\%}}{J_{c,0\%}}(H)$ value in $H$ < 0.1 T regime, see figs.6(b), 6(d)). However at 44 K, in the high $H$ regime (> 0.1 T, viz., strong intervortex interaction regime) we observe that although $\frac{J_{c,0.05\%}}{J_{c,0\%}}(H) > 1$, it is still lower than the $\frac{J_{c,0.05\%}}{J_{c,0\%}}$ value at $H$ < 0.1 T (see fig.6(b) and fig.6(d)). This feature at 44 K arises, because the net vortex state becomes effectively more rigid with increase in $H$ (> 0.1 T), as the enhancement in vortex state rigidity at higher $H$ is a much larger than the thermally induced softening of the vortex state at 44 K. Thus, a net higher vortex rigidity found at 44 K in the $H$ > 0.1 T regime, results in the progressive weakening of vortex pinning with increase in $H$. Consequently the, $\frac{J_{c,0.05\%}}{J_{c,0\%}}(H)$ or $\frac{J_{c,2\%}}{J_{c,0\%}}(H)$, at 44 K shows a decreasing trend with increase in $H$ > 0.1 T. At 77 K, however, the significantly enhanced thermal softening effect completely dominates over the increase in vortex state rigidity caused by the increase in $H$. Hence, even at high $H$ (> 0.1 T) a net softened vortex state at 77 K accommodates the pinning centers of $Co_2C$ NP's well. In contrast, at very low $T$ of 5 K due to very weak thermal smearing out of pinning effects, the pinning strength remains at a very high value at 5 K. At 5 K the thermal vortex softening effects are negligible. Hence at 5K in the strong pinning regime produced by the $Co_2C$ NP's, the pinning completely dominates over the intervortex interactions effects. Therefore the vortex state at 5 K shows a saturated strong pinning regime over the entire $H$ range and hence $\frac{J_{c,0.05\%}}{J_{c,0\%}}(H)$ or $\frac{J_{c,2\%}}{J_{c,0\%}}(H)$ behavior is almost flat. We see upto 1 T at 5 K the rigidity of the vortex state isn't strong enough to overcome the pinning effects and thereby cause any decrease of $\frac{J_{c,0.05\%}}{J_{c,0\%}}(H)$ or $\frac{J_{c,2\%}}{J_{c,0\%}}(H)$. Thus the 5K, 44 K and 77 K data in figs.6(b) and 6(d) illustrates the competition and interplay between different effects in our nanocomposite, viz., (i) thermal fluctuation induced smearing (weakening) of the pinning potential, (ii) thermal fluctuation induced softening of the vortex state by increasing temperature and (iii) magnetic field induced enhancement in rigidity of the vortex state. Since we observe that pinning is not smeared out at and strong pinning characteristics are retained at 77 K pinning properties in our $Co_2C$ – Bi-2223 nanocomposite, therefore get an estimate of the lower bound on the depth of the pinning potential generated by $Co_2C$ NP's in Bi-2223 to be atleast of $77k_B$. This is a potentially useful property as it suggests that its possible to achieve and retain sufficiently high $J_c$ in the high T regime (near $T_c$ or atleast upto 77 K) in this Bi-2223-$Co_2C$ nanocomposite. From the table also we see that, w.r.t. the pristine (0%) pellet, while there is a significant increase in $J_c$ for the nanocomposite pellet, however between 0.05% and 2% pellet there isn't a proportionate significant change in $J_c$. We believe, this suggests predominance of strong individual vortex-pin interactions in this nanocomposite pellet, namely, the interaction generated by the local pair breaker ($Co_2C$ NP's) sites with vortices in their neighbourhood, and these dominate over collective interaction effects. This is supported by our observation (see table-1) of nearly similar low field range (upto $H^*$) over which strong vortex pinning (SVP) regime ($J_c(H)\sim$ constant) exists for the 0.05% and 2% nanocomposites (figs.6(a) and 6(c)) and nearly similar values of the $J_c$ in the SVP regime. The high critical current density achieved in the $Co_2C$-Bi-2223 nanoparticle composites, especially, in the relatively high *T* regime (77 K range), is potentially useful for high current applications. However, for future application point of view one also needs to find ways to avoid the large sized clustering tendency of magnetic $Co_2C$ NP's which is primarily responsible for loss of superconducting fraction in the nanocomposite.

**TABLE-1**

➕ **Table Summarizing the Properties of the $Co_2C$-Bi-2223 nanocomposites:**

| Sample | $T_c$ (K) | $H_{irr}$ (T) at 77K | $J_c$ @ 5K, 0.01T (Am$^{-2}$) | $J_c$ @77K, 0.01T (Am$^{-2}$) |
|---|---|---|---|---|
| Bi-2223 with 0% added $Co_2C$ | 108 | 0.3 | $2.2\times10^9$ | $9.03\times10^7$ |
| Bi-2223 with 0.05% added $Co_2C$ | 108 | 0.9 | $2.53\times10^{10}$ | $6.05\times10^9$ |
| Bi-2223 with 2% added $Co_2C$ | 108 | 0.91 | $2.30\times10^{10}$ | $6.68\times10^9$ |

**Conclusion:**

We have investigated the pinning properties of composites made from NPs of Bi-2223 high $T_c$ superconductor mixed with NPs of FM $Co_2C$. We show that the presence of FM in the mechanically pressed composites does not destroy the SC fraction present in the pellet. The surviving superconducting phase exhibits robust bulk superconductivity with negligible degradation of its $T_c$. These SC-FM composites exhibit significant enhancement in pinning strength. From the application point of view, a unique and lucrative feature of these composites is our finding that there is a significantly enhanced critical current regime in the high $T$ regime. Also, the irreversibility field ($H_{irr}$) field has increased by a factor of 3 which implies the critical current retains up to a very high applied magnetic field and this property can be very useful for high field application of the admixed-superconductor. We are exploring the possibility of developing $Co_2C$ NP based HTSC - high $J_c$ wires. In this context, it is necessary to assess the effects of strain and thermal treatment on the pinning properties on $Co_2C$ NP's which are inevitable effects present during wire fabrication process. These are the directions of our ongoing efforts as well as trying to develop a better understanding of the microscopic pinning mechanism by $Co_2C$ NP's in HTSC materials.

**Supplementary Material:** In the supplementary material we have provided EDS compositional analysis of the pristine Bi-2223 pellet, which confirms that the pristine sample is in the pure BSCCO-2223 phase. Also, fig.-2 shows variation of magnetization with temperature for 0%, 0.05%, 2% and 10% Bi-2223 admixed pellet, which tells that there is decrease in superconducting fraction with increase in $Co_2C$ admixture percentage.

**Acknowledgements**: S S Banerjee thanks DST – SUPRA and AMT Government of India and IIT Kanpur for funding support. Md. Arif Ali thanks IIT Kanpur for funding support. Sourav M Karan thanks UGC for the funding support.

**Note**: Md. Arif Ali, Sourav M Karan are equal contributors in the present work.

**References :**

1  G. Blatter, M. V. Feigel'man, V. B. Geshkenbein, A. I. Larkin, and V. M. Vinokur, "Vortices in high-temperature superconductors," Reviews of Modern Physics **66** (4), 1125-1388 (1994).
2  Michael Tinkham, *Introduction to superconductivity*. (Courier Corporation, 2004); B. Bag, D. J. Sivananda, P. Mandal, S. S. Banerjee, A. K. Sood, and A. K. Grover, "Vortex depinning as a nonequilibrium phase transition phenomenon: Scaling of current-voltage curves near the low and the high critical-current states in 2H-$NbS_2$ single crystals," Phys Rev B **97** (13) (2018).

3   C. Reichhardt and C. J. Olson Reichhardt, "Depinning and nonequilibrium dynamic phases of particle assemblies driven over random and ordered substrates: a review," Reports on Progress in Physics **80** (2), 026501 (2016).
4   David Larbalestier, Alex Gurevich, D. Matthew Feldmann, and Anatoly Polyanskii, "High-$T_c$ superconducting materials for electric power applications," Nature **414** (6861), 368-377 (2001).
5   Andreas Glatz, Ivan A. Sadovskyy, Ulrich Welp, Wai-Kwong Kwok, and George W. Crabtree, "The Quest for High Critical Current in Applied High-Temperature Superconductors," Journal of Superconductivity and Novel Magnetism **33** (1), 127-141 (2020).
6   H. Hilgenkamp and J. Mannhart, "Grain boundaries in high-$T_c$ superconductors," Reviews of Modern Physics **74** (2), 485-549 (2002).
7   K. Harada, O. Kamimura, H. Kasai, T. Matsuda, A. Tonomura, and V. V. Moshchalkov, "Direct Observation of Vortex Dynamics in Superconducting Films with Regular Arrays of Defects," Science **274** (5290), 1167-1170 (1996).
8   R. Abd-Shukor and A. N. Jannah, "Advances in superconductivity and $Co_3O_4$ nanoparticles as flux pinning center in (Bi, Pb)-2223/Ag superconductor tapes," AIP Conference Proceedings **1877** (1), 020003 (2017).
9   Masashi Miura, Go Tsuchiya, Takumu Harada, Keita Sakuma, Hodaka Kurokawa, Naoto Sekiya, Yasuyuki Kato, Ryuji Yoshida, Takeharu Kato, Koichi Nakaoka, Teruo Izumi, Fuyuki Nabeshima, Atsutaka Maeda, Tatsumori Okada, Satoshi Awaji, Leonardo Civale, and Boris Maiorov, "Thermodynamic approach for enhancing superconducting critical current performance," NPG Asia Materials **14** (1), 85 (2022).
10  D. Dew-Hughes, "The critical current of superconductors: an historical review," Low Temperature Physics **27** (9), 713-722 (2001).
11  A. Galluzzi, K. Buchkov, E. Nazarova, V. Tomov, G. Grimaldi, A. Leo, S. Pace, and M. Polichetti, "Pinning energy and anisotropy properties of a Fe(Se, Te) iron based superconductor," Nanotechnology **30** (25), 254001 (2019).
12  S Senoussi, "Review of the critical current densities and magnetic irreversibilities in high Tc superconductors," Journal de physique III **2** (7), 1041-1257 (1992).
13  J. Jiang, X. Y. Cai, A. A. Polyanskii, L. A. Schwartzkopf, D. C. Larbalestier, R. D. Parrella, Q. Li, M. W. Rupich, and G. N. Riley, "Through-process study of factors controlling the critical current density of Ag-sheathed $(Bi,Pb)_2Sr_2Ca_2Cu_3O_x$ tapes," Superconductor Science and Technology **14** (8), 548-556 (2001).
14  Jijie Huang and Haiyan Wang, "Effective magnetic pinning schemes for enhanced superconducting property in high temperature superconductor $YBa_2Cu_3O_{7-x}$: a review," Superconductor Science and Technology **30** (11), 114004 (2017).
15  S. H. Pan, E. W. Hudson, K. M. Lang, H. Eisaki, S. Uchida, and J. C. Davis, "Imaging the effects of individual zinc impurity atoms on superconductivity in $Bi_2Sr_2CaCu_2O_{8+\delta}$," Nature **403** (6771), 746-750 (2000);    S. H. Pan, J. P. O'Neal, R. L. Badzey, C. Chamon, H. Ding, J. R. Engelbrecht, Z. Wang, H. Eisaki, S. Uchida, A. K. Gupta, K. W. Ng, E. W. Hudson, K. M. Lang, and J. C. Davis, "Microscopic electronic inhomogeneity in the high-Tc superconductor $Bi_2Sr_2CaCu_2O_{8+x}$," Nature **413** (6853), 282-285 (2001);    P. H. Kes, "Flux pinning and creep in high temperature superconductors," Physica C: Superconductivity **185-189**, 288-291 (1991);    A. Díaz, L. Mechin, P. Berghuis, and J. E. Evetts, "Evidence for Vortex Pinning by Dislocations in $YBa_2Cu_3O_{7-delta}$ Low-Angle Grain Boundaries," Physical Review Letters **80** (17), 3855-3858 (1998);    B. Dam, J. M. Huijbregtse, F. C. Klaassen, R. C. F. van der Geest, G. Doornbos, J. H. Rector, A. M. Testa, S. Freisem, J. C. Martinez, B. Stäuble-Pümpin, and R. Griessen, "Origin of high critical currents in $YBa_2Cu_3O_{7-\delta}$ superconducting thin films," Nature **399** (6735), 439-442 (1999).
16  L. Piraux and X. Hallet, "Artificial vortex pinning arrays in superconducting films deposited on highly ordered anodic alumina templates," Nanotechnology **23** (35), 355301 (2012).
17  T. Haugan, P. N. Barnes, R. Wheeler, F. Meisenkothen, and M. Sumption, "Addition of nanoparticle dispersions to enhance flux pinning of the $YBa_2Cu_3O_{7-x}$ superconductor," Nature **430** (7002), 867-870 (2004);    V. G. Prabitha, A. Biju, R. G. Abhilash Kumar, P. M. Sarun, R. P. Aloysius, and U. Syamaprasad, "Effect of Sm addition on (Bi,Pb)-2212 superconductor," Physica C: Superconductivity **433** (1), 28-36 (2005);    J. L. MacManus-Driscoll, S. R.

| | |
|---|---|
| | Foltyn, Q. X. Jia, H. Wang, A. Serquis, L. Civale, B. Maiorov, M. E. Hawley, M. P. Maley, and D. E. Peterson, "Strongly enhanced current densities in superconducting coated conductors of $YBa_2Cu_3O_{7-x}$ + $BaZrO_3$," Nature Materials **3** (7), 439-443 (2004). |
| 18 | R. Abd-Shukor and W. Kong, "Magnetic field dependent critical current density of Bi-Sr-Ca-Cu-O superconductor in bulk and tape form with addition of $Fe_3O_4$ magnetic nanoparticles," J Appl Phys **105** (7), 07E311 (2009). |
| 19 | Gang Xiao, Marta Z. Cieplak, J. Q. Xiao, and C. L. Chien, "Magnetic pair-breaking effects: Moment formation and critical doping level in superconducting $La_{1.85}Sr_{0.15}Cu_{1-x}A_xO_4$ systems (A=Fe,Co,Ni,Zn,Ga,Al)," Phys Rev B **42** (13), 8752-8755 (1990). |
| 20 | M. K. Yu and J. P. Franck, "Comparison of the low-temperature specific heat of Fe- and Co-doped $Bi_{1.8}Pb_{0.2}Sr_2Ca(Cu_{1-x}M_x)_2O_8$ (M=Fe or Co): Anomalously enhanced electronic contribution due to Fe doping," Phys Rev B **53** (13), 8651-8657 (1996). |
| 21 | J. Li, Y. F. Guo, S. B. Zhang, J. Yuan, Y. Tsujimoto, X. Wang, C. I. Sathish, Y. Sun, S. Yu, W. Yi, K. Yamaura, E. Takayama-Muromachiu, Y. Shirako, M. Akaogi, and H. Kontani, "Superconductivity suppression of $Ba_{0.5}K_{0.5}Fe_{2-2x}M_{2x}As_2$ single crystals by substitution of transition metal (M = Mn, Ru, Co, Ni, Cu, and Zn)," Phys Rev B **85** (21), 214509 (2012). |
| 22 | H. B. Lee, G. C. Kim, Young Jin Shon, Dongjin Kim, and Y. C. Kim, "Flux-pinning behaviors and mechanism according to dopant level in (Fe, Ti) particle-doped $MgB_2$ superconductor," Scientific Reports **11** (1), 10564 (2021); I. F. Lyuksyutov and D. G. Naugle, "FROZEN FLUX SUPERCONDUCTORS," Modern Physics Letters B **13** (15), 491-497 (1999). |
| 23 | Changwei Zou, Zhenqi Hao, Haiwei Li, Xintong Li, Shusen Ye, Li Yu, Chengtian Lin, and Yayu Wang, "Effect of Structural Supermodulation on Superconductivity in Trilayer Cuprate $Bi_2Sr_2Ca_2Cu_3O_{10+delta}$," Physical Review Letters **124** (4), 047003 (2020). |
| 24 | D. C. Larbalestier, J. Jiang, U. P. Trociewitz, F. Kametani, C. Scheuerlein, M. Dalban-Canassy, M. Matras, P. Chen, N. C. Craig, P. J. Lee, and E. E. Hellstrom, "Isotropic round-wire multifilament cuprate superconductor for generation of magnetic fields above 30 T," Nature Materials **13** (4), 375-381 (2014). |
| 25 | F. Kametani, J. Jiang, M. Matras, D. Abraimov, E. E. Hellstrom, and D. C. Larbalestier, "Comparison of growth texture in round Bi2212 and flat Bi2223 wires and its relation to high critical current density development," Scientific Reports **5** (1), 8285 (2015). |
| 26 | Chao Yao and Yanwei Ma, "Superconducting materials: Challenges and opportunities for large-scale applications," iScience **24** (6), 102541 (2021). |
| 27 | Hesam Fallah-Arani, Saeid Baghshahi, Arman Sedghi, and Nastaran Riahi-Noori, "Enhancement in the performance of BSCCO (Bi-2223) superconductor with functionalized $TiO_2$ nanorod additive," Ceramics International **45** (17, Part A), 21878-21886 (2019). |
| 28 | Ali Aftabi and Morteza Mozaffari, "Intergranular Coupling, Critical Current Density, and Phase Formation Enhancement of Polycrystalline $Bi_{1.6}Pb_{0.4}Sr_2Ca_2Cu_3O_{10-y}$ Superconductors by α-$Al_2O_3$ Nanoparticle Addition," Journal of Superconductivity and Novel Magnetism **28** (8), 2337-2343 (2015). |
| 29 | Ali Aftabi and Morteza Mozaffari, "Fluctuation induced conductivity and pseudogap state studies of $Bi_{1.6}Pb_{0.4}Sr_2Ca_2Cu_3O_{10+\delta}$ superconductor added with ZnO nanoparticles," Scientific Reports **11** (1), 4341 (2021); Xiao-Jia Chen, Viktor V. Struzhkin, Yong Yu, Alexander F. Goncharov, Cheng-Tian Lin, Ho-kwang Mao, and Russell J. Hemley, "Enhancement of superconductivity by pressure-driven competition in electronic order," Nature **466** (7309), 950-953 (2010). |
| 30 | V. G. Harris, Y. Chen, A. Yang, S. Yoon, Z. Chen, A. L. Geiler, J. Gao, C. N. Chinnasamy, L. H. Lewis, C. Vittoria, E. E. Carpenter, K. J. Carroll, R. Goswami, M. A. Willard, L. Kurihara, M. Gjoka, and O. Kalogirou, "High coercivity cobalt carbide nanoparticles processed via polyol reaction: a new permanent magnet material," Journal of Physics D: Applied Physics **43** (16), 165003 (2010). |
| 31 | Zhanbing He, Jean-Luc Maurice, Aurélien Gohier, Chang Seok Lee, Didier Pribat, and Costel Sorin Cojocaru, "Iron Catalysts for the Growth of Carbon Nanofibers: Fe, $Fe_3C$ or Both?," Chemistry of Materials **23** (24), 5379-5387 (2011). |

32   J. J. Host, J. A. Block, K. Parvin, V. P. Dravid, J. L. Alpers, T. Sezen, and R. LaDuca, "Effect of annealing on the structure and magnetic properties of graphite encapsulated nickel and cobalt nanocrystals," J Appl Phys **83** (2), 793-801 (1998).

33   Kyler J. Carroll, Zachary J. Huba, Steven R. Spurgeon, Meichun Qian, Shiv N. Khanna, Daniel M. Hudgins, Mitra L. Taheri, and Everett E. Carpenter, "Magnetic properties of $Co_2C$ and $Co_3C$ nanoparticles and their assemblies," Applied Physics Letters **101** (1), 012409 (2012).

34   Xianfeng Shen, Tianfu Zhang, HaiYun Suo, Lai Yan, Lichun Huang, Chenwei Ma, Linge Li, Xiaodong Wen, Yongwang Li, and Yong Yang, "A facile one-pot method for synthesis of single phase $Co_2C$ with magnetic properties," Materials Letters **271**, 127783 (2020).

35   Xianfeng Shen, Chenwei Ma, HaiYun Suo, Tianfu Zhang, Lai Yan, Lichun Huang, Jianqiang Zhou, Xiaodong Wen, Yongwang Li, and Yong Yang, "Wet-chemistry approach for the synthesis of single phase ferromagnetic $Co_3C$ nanoparticle," Nano Select **2** (7), 1368-1371 (2021).

36   Nirmal Roy, Arpita Sen, Prasenjit Sen, and S. S. Banerjee, "Localized spin waves at low temperatures in a cobalt carbide nanocomposite," J Appl Phys **127** (12), 124301 (2020).

37   Nirmal Roy, Md Arif Ali, Arpita Sen, D. T. Adroja, Prasenjit Sen, and S. S. Banerjee, "Exploring a low temperature glassy state, exchange bias effect, and high magnetic anisotropy in $Co_2C$ nanoparticles," Journal of Physics: Condensed Matter **33** (37), 375804 (2021).

38   Mannan Ali, Patrick Adie, Christopher H. Marrows, Denis Greig, Bryan J. Hickey, and Robert L. Stamps, "Exchange bias using a spin glass," Nature Materials **6** (1), 70-75 (2007).

39   W. H. Meiklejohn and C. P. Bean, "New Magnetic Anisotropy," Physical Review **102** (5), 1413-1414 (1956).

40   N. H. Mohammed, Ramadan Awad, A. I. Abou-Aly, I. H. Ibrahim, and M. S. Hassan, "Optimizing the Preparation Conditions of Bi-2223 Superconducting Phase Using PbO and $PbO_2$," **Vol.03No.04**, 10 (2012).

41   Charles P Poole, Horacio A Farach, and Richard J Creswick, Superconductivity. (Academic press, 2013).

42   Shigeyuki Ishida, Daniel Kagerbauer, Sigrid Holleis, Kazuki Iida, Koji Munakata, Akiko Nakao, Akira Iyo, Hiraku Ogino, Kenji Kawashima, Michael Eisterer, and Hiroshi Eisaki, "Superconductivity-driven ferromagnetism and spin manipulation using vortices in the magnetic superconductor $EuRbFe_4As_4$," Proceedings of the National Academy of Sciences **118** (37), e2101101118 (2021).

43   Charles P. Bean, "Magnetization of High-Field Superconductors," Reviews of Modern Physics **36** (1), 31-39 (1964).

44   C. P. Bean, "Magnetization of Hard Superconductors," Physical Review Letters **8** (6), 250-253 (1962).

45   Md Arif Ali and S. S. Banerjee, "Coexistence of different pinning mechanisms in Bi-2223 superconductor and its implications for using the material for high current applications," J Appl Phys **131** (24), 243901 (2022).

46   A. I. Larkin and Yu N. Ovchinnikov, "Pinning in type II superconductors," Journal of Low Temperature Physics **34** (3), 409-428 (1979).

47   C. J. van der Beek, M. Konczykowski, A. Abal'oshev, I. Abal'osheva, P. Gierlowski, S. J. Lewandowski, M. V. Indenbom, and S. Barbanera, "Strong pinning in high-temperature superconducting films," Phys Rev B **66** (2), 024523 (2002).

48   N. Haberkorn, S. Suárez, S. L. Bud'ko, and P. C. Canfield, "Strong pinning and slow flux creep relaxation in Co-doped $CaFe_2As_2$ single crystals," Solid State Communications **318**, 113963 (2020).

49   Lina Sang, Pankaj Maheswari, Zhenwei Yu, Frank F. Yun, Yibing Zhang, Shixue Dou, Chuanbing Cai, V. P. S. Awana, and Xiaolin Wang, "Point defect induced giant enhancement of flux pinning in Co-doped $FeSe_{0.5}Te_{0.5}$ superconducting single crystals," AIP Advances **7** (11), 115016 (2017).

50   A. I. Kosse, A. Y. Prokhorov, V. A. Khokhlov, G. G. Levchenko, A. V. Semenov, D. G. Kovalchuk, M. P. Chernomorets, and P. N. Mikheenko, "Measurements of the magnetic field and temperature dependences of the critical current in YBCO films and procedures for an appropriate theoretical model selection," Superconductor Science & Technology **21** (7) (2008).

51  N. Haberkorn, J. Kim, K. Gofryk, F. Ronning, A. S. Sefat, L. Fang, U. Welp, W. K. Kwok, and L. Civale, "Enhancement of the critical current density by increasing the collective pinning energy in heavy ion irradiated Co-doped BaFe$_2$As$_2$ single crystals," Superconductor Science& Technology **28** (5) (2015).
52  Y. Oner and C. Boyraz, "Critical current density and flux pinning in BaFe$_{1.9}$Pt$_{0.1}$As$_2$ and La doped Ba$_{0.9}$5La$_{0.05}$Fe$_{1.9}$Pt$_{0.1}$As$_2$ polycrystals," Journal of Physics-Condensed Matter **31** (15) (2019).

# Impact of Co$_2$C Nanoparticles on Enhancing the Critical Current Density of Bi-2223 Superconductor

Md. Arif Ali[1], Sourav M Karan[1], Nirmal Roy[1], S. S. Banerjee[1,*]

[1]*Department of Physics, Indian Institute of Technology Kanpur, Kanpur, Uttar Pradesh 208016, India*

# Supplementary section-I:

> Compositional Analysis of BSCCO-2223 from EDS:

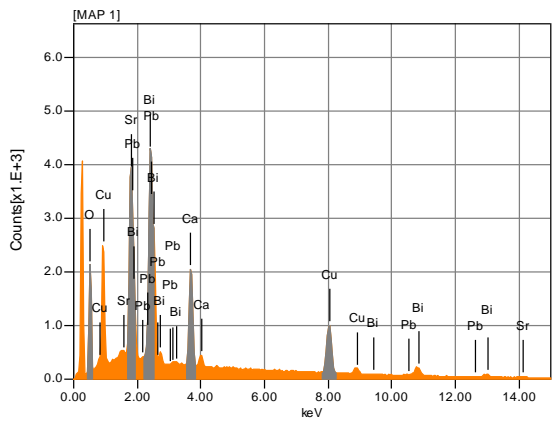

| Chemical Formula | Mass Percentage | Atomic Percentage | Line |
|---|---|---|---|
| Bi | 40.98 | 11.65 | M |
| Pb | 3.83 | 1.10 | M |
| Sr | 18.36 | 12.45 | L |
| Cu | 16.41 | 15.35 | K |
| Ca | 7.34 | 10.89 | K |
| O | 13.07 | 48.55 | K |

- Molar Ratio of the elements of BSCCO-2223 Powder,

Bi : Pb : Sr : Ca : Cu = 11.65 : 1.10 : 12.45 : 15.35 : 10.89
= 2.277 : 0.215 : 2.433 : 2.128 : 3

- Comparison of Superconducting Magnetization ($M_{sc}$) vs T for 0%, 0.05%, 2%, and 10% Bi-2223 Admixed Sample (similar to that shown in fig.2(c) of the main paper). We see for all the pellets, the $T_c$ of the sample remains unchanged. At T well below $T_c$ we clearly see a progressive decrease in the $M_{sc}$ value with increase in the $Co_2C$ percentage.

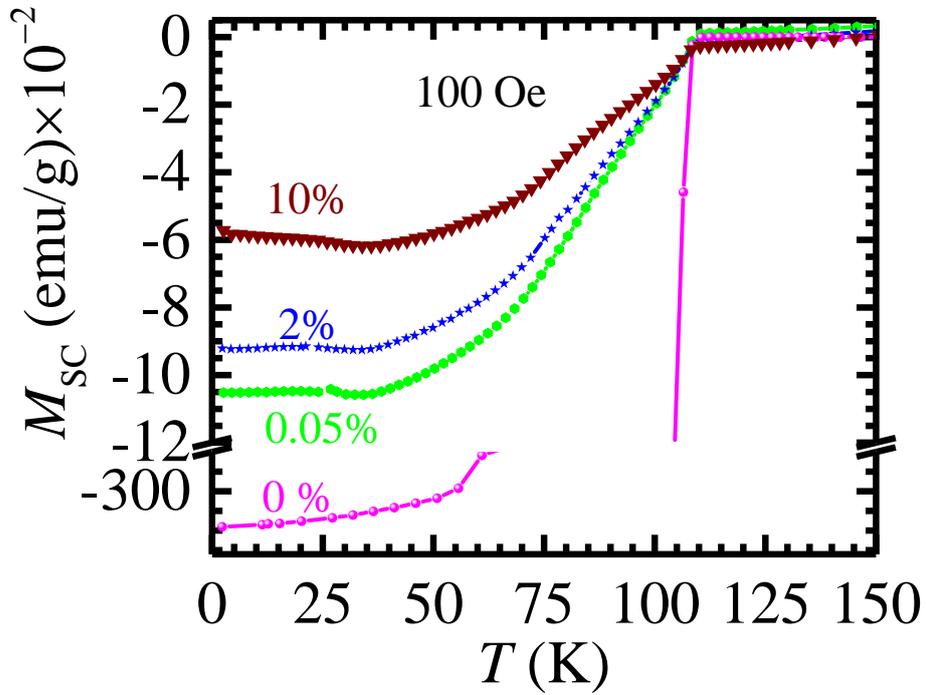

Fig.2: Magnetization vs Temperature for 0%, 0.05%, 2% and 10% Bi-2223 admixed pellet, which shows the decrease in superconducting fraction with increase in $Co_2C$ admixture percentage.